\documentclass[12pt]{article}
\pdfoutput=1

\usepackage[utf8]{inputenc}
\usepackage{CJKutf8}
\usepackage{setspace}
\usepackage{makecell}
\usepackage{array}
\usepackage{multirow}
\usepackage{tabularx}
\usepackage{graphicx}
\usepackage{subcaption}
\usepackage{geometry}
\usepackage{indentfirst}
\usepackage{caption}
\usepackage{booktabs}
\usepackage{amsmath}
\usepackage{amssymb}
\usepackage{xurl}
\usepackage[style=numeric]{biblatex}
\usepackage{titlesec}
\usepackage{fancyhdr}
\usepackage{hyperref}
\usepackage{times}

\savegeometry{global}
\newcommand{\keywords}[1]{\textbf{关键词：} #1}

\newcommand{\subsubsubsection}[1]{\paragraph{#1}\mbox{}\\}
\newcommand{\zihao}[1]{%
  \ifnum#1=6 \fontsize{9pt}{0pt}\selectfont\fi
  \ifnum#1=5 \fontsize{10pt}{12pt}\selectfont\fi
}

\begin{document}
\begin{CJK}{UTF8}{gbsn}
\title{产业结构转型升级与新质生产力发展\\——基于2003—2022年中国省级面板数据的实证分析}
\author{金阳\\Solar Jin}
\date{}
\maketitle

\begin{abstract}
    加快产业结构深度转型升级，加快形成新质生产力是我国实现伟大复兴中国梦的重要组成部分。经过40多年的高速发展，我国的发展进入“新常态”，发展新质生产力刻不容缓。本文梳理了中国产业结构演变历程，论证了中国进行新一轮产业深度转型的必要性，并利用多种方法探讨了产业结构转型升级对新质生产力水平的影响。研究结果发现：(1)产业深度转型升级能够显著促进新质生产力水平发展，但存在较为明显的地区差异。(2)新质生产力水平的提升的核心标志是全要素生产率的提升。更进一步，本文归纳总结了过往的产业发展历程和发展时面临过的挑战，并结合实证研究结果分析讨论了产业结构深度转型促进新质生产力发展可能面临的挑战。
\end{abstract}

\section{引言}
高质量发展是实现中华民族伟大复兴中国梦的重要实现步骤，大国崛起离不开高水平发展的关键支撑。改革开放以来，我国的发展取得了历史性的成就，当前，我国经济总量已跃居世界第二，制造业和货物贸易规模稳居世界第一，为世界发展作出了积极贡献。特别是党的十八大以来，我国全面贯彻新发展理念，发展的核心从高速增长转向高质量发展。2023年习总书记首次提出“加快形成新质生产力”的目标。我国正处于从“量的扩张”向“质的提升”转型的关键时期，优化产业结构已然成为了推动新质生产力发展的重要战略举措。2024年习总书记在二十届中央政治局第十一次集体学习上强调，新质生产力由技术革命性突破、生产要素创新性配置、产业深度转型升级而催生，以劳动者、劳动资料、劳动对象及其优化组合的跃升为基本内涵，以全要素生产率大幅提升为核心标志，特点是创新，关键在质优，本质是先进生产力。
\par
2012年以来，中国就开始了产业升级与结构优化的步伐。2012年的国务院政府工作报告中写到，“加快转变经济发展方式，提高发展的协调性和产业的竞争力。我们坚持有扶有控，促进结构调整和优化升级，增强发展后劲”。随后我国产业加速转型，几年内出台了数十条具体的产业结构优化政策。
\par
基于此，本文结合产业结构转型和工业结构升级的双重属性\cite{ZJSY201801014}梳理了新中国产业结构演变的历程，论证了中国新一轮产业结构深度转型升级对新质生产力发展的必要性，并以2003—2022年省级层面的各项数据构造的新质生产力指数，利用各省的经济数据，证明了产业深度转型对中国新质生产力发展的推动作用。实证方面的研究发现：(1)产业深度转型升级能够显著促进新质生产力水平发展，但存在较为明显的地区差异。(2)新质生产力水平的提升的核心标志是全要素生产率的提升。
\par
另外，高质量发展发展之路充满挑战。我国进入高质量发展新阶段，需要总结我国以及世界各地产业转型的优秀经验教训，充分利用我国的制度优势，提高我国的产业的转型效率与竞争力。世界正处于百年未有之大变局，尤其是在技术进步和全球化加速的背景下，我国产业转型不仅是高质量发展的关键一步，也是各国在全球价值链中重新定位的过程。为此，本文归纳总结了世界各国过往的产业发展历程和发展时面临过的挑战，并结合实证研究结果、现今的全球发展格局分析展望了我国产业结构深度转型促进新质生产力发展的未来发展道路。
\par
全文的结构安排如下：第二部分为中国产业结构演变历程和新一轮产业结构深度转型升级的必要性；第三部分为理论假说；第四部分是实证研究；第五部分论述产业转型促进新质生产力发展可能面临的挑战；第六部分为结论。

\section{中国产业结构演变历程和新一轮产业深度转型的必要性}
\subsection{中国产业结构演变的历程}
\subsubsection{第一阶段：纠正失衡和均衡发展的产业结构}
    1978年，我国产业结构呈现“二一三”格局，三次产业比例为27.7：47.7：24.6。在1978年十一届三中全会后，中国的经济进入了改革开放的新阶段。改革初期，产业结构失调成为了经济发展的突出问题，因此调整产业结构，解决经济中的不平衡问题成为改革开放初期的核心任务。1979年至1980年底的第一阶段，改革重点是调整国民经济中的农、轻、重产业关系，特别是通过改善城乡居民收入和消费结构，为经济健康发展打下基础。通过实施家庭联产承包责任制，农业生产力得到了释放，农民收入增加，农业生产效率提升。而在城市，改革政策促进了轻工业和重工业的协调发展，推动了城乡收入的平衡和经济的整体调整。此时，国家不仅关注农业和工业的平衡，还通过控制消费和加大投资力度，优化了各产业的比重关系。
\par
    改革的初步成效在于，第一产业的比重显著下降，第二产业和第三产业得到了稳步发展。特别是第三产业，随着人民生活水平的提高，逐渐成为经济增长的亮点。1981年，第一产业在GDP中的比重已从33.4\%降至19.7\%，而第三产业的比重则从21.8\%上升至32.8\%，显示出服务业逐步在经济中占据了更加重要的位置。与此同时，就业结构也发生了变化，农民逐渐进入城市，进入第二、第三产业，从而推动了社会就业水平的提高。

\subsubsection{第二阶段：重化工业重启下的产业结构}
    1985年第三产业规模首次超过第一产业，三次产业比例实现“二一三”向“二三一”的重大转变，三次产业比例调整为27.9：42.7：29.4。在这一阶段，中国经济经历了重要的转型和调整。随着市场经济体制的初步建立，经济运行模式也发生了转变。过去的供给约束型经济逐渐转向需求约束型经济，经济增长动力的结构发生了变化。在这一时期，政府大力推动基本建设投资。这些建设推动了重化工业的再次崛起，并加速了工业结构的调整与升级。
\par
    1998年以后，中国出口迎来了高速增长期，尤其是在加入WTO后，中国利用国际市场和资源，加快了全球分工的步伐，成为世界制造业的中心。中国的商品出口在全球市场的份额大幅提高，2007年超过美国，2009年超过德国，成为全球最大的商品出口国。这种出口增长不仅体现了中国制造业的崛起，也反映了中国在全球产业链中的地位变化。中国不仅是世界的“制造工厂”，还迅速成为高技术产品和信息通信技术产品的全球制造中心。到2010年，中国的高技术产品出口占全球市场的22.2\%。这标志着中国在全球价值链中的地位从传统的低技术制造转向高技术、附加值较高的产业领域。
\par
    经济增长动力的变化促使产业结构迅速变化。第二产业继续占据中国经济的主导地位，1998至2012年间，其产值占比稳定在45\%左右，而第三产业的比重则从36.2\%上升到44.6\%。与此同时，第一产业的比重从17.6\%降至10.1\%。在工业内部结构上，重化工业的比重大幅上升，这些行业成为经济增长的主要驱动力。然而，这种增长方式也带来了结构性问题，如高耗能重化工业的比重过高，工业能耗的70\%以上来自于六大高耗能行业，这也表明中国在这一阶段面临着产业结构优化的压力。

\subsubsection{第三阶段：经济新常态下的产业结构}
    2012年第三产业规模再次超过第二产业，成为推动国民经济发展的主导产业，三次产业结构实现“二三一”向“三二一”的历史性转变，三次产业比例调整为9.1：45.4：45.5。同时，经济发展迈入了“三期叠加”阶段，实现“双中高”目标是在“五大发展理念”指引下的重要目标，也是提质增效的经济保障，而供给侧结构性改革的核心，即产业升级与结构转型则是解决矛盾的必要手段。面对这些问题，政府提出了供给侧结构性改革的核心目标，即通过调整产业结构、提升技术水平、推动绿色和高端制造业的发展，解决经济发展中的不平衡和不充分问题。这一阶段的政策重点是加强创新驱动，推动传统制造业向智能化、高端化转型，并加大对战略性新兴产业的支持。此外，实施“绿色发展”和“智能制造”战略，推进传统产业的技术升级和资源高效利用。2013年，中国提出了“一带一路”倡议，这一战略不仅是中国外交政策的重要组成部分，也为中国产业结构转型提供了新的机遇。通过对外投资和基础设施建设，特别是在高铁、核电、港口等领域的合作，中国在国际市场上拓展了新的增长点。特别是中国制造业和高端装备的“走出去”战略，有效缓解了国内一些产业的过剩问题，并带动了相关技术、服务和资本的输出。“一带一路”倡议促进了中国产业在全球价值链中的升级，为国内的产业结构调整提供了外部动力，推动了产业结构向更高端、更国际化的方向发展。
\par
    进入2016年后，中国政府明确提出供给侧结构性改革的战略方向，强调通过提升供给质量来满足日益多样化和高层次的消费需求。在这一阶段，产业结构的优化更加注重资源配置效率和创新驱动，推动制造业与服务业的深度融合，尤其是在高端制造、数字经济和现代服务业领域的融合创新。中国加强了对战略性新兴产业的支持，尤其是在互联网、大数据、云计算等领域的发展，推动了产业的智能化转型。同时，产业结构调整更加注重绿色低碳发展，推进清洁能源、环保技术等绿色产业的布局，实施“绿色制造”标准和政策，促进了资源节约与环境保护。
\par
    从2020年开始，中国进入了经济高质量发展的新时代。疫情后的全球经济环境复杂多变，推动中国更加注重经济的内生增长动力，并加大科技创新和数字化转型的力度。中国加快了制造业的智能化升级，推动“工业4.0”向“工业5.0”过渡，加强了制造业与科技创新的深度融合。同时，中国提出了“创新驱动发展”的战略，强调科技创新是提升产业竞争力的核心动力，进一步推动了信息技术与传统产业的深度融合。在这一时期，人工智能、5G通讯、大数据、区块链等技术迅速发展，成为新经济的重要支撑。这些新兴技术不仅推动了新兴产业的发展，也带动了传统产业的数字化转型。同时，产业向更高附加值、高技术含量的方向发展，推动了制造业的质量提升和创新能力的增强。近年来，随着全球对气候变化问题的关注加剧，中国提出了“双碳”目标，即到2030年碳达峰、2060年碳中和，这一目标的提出再次为产业结构调整提供了重要的政策导向。绿色低碳产业成为新的发展方向，特别是在新能源、绿色金融、绿色制造等领域，相关政策不断出台，推动产业结构向绿色、低碳、高效转型。中国的新能源产业快速发展，并且在全球市场中占据了领先地位。同时，传统能源行业也在进行转型升级。这一过程中，技术创新、产业协同和政策支持是产业升级的关键因素。

\subsection{新一轮产业结构深度转型升级的必要性}
    党的二十届三中全会审议通过的《中共中央关于进一步全面深化改革、推进中国式现代化的决定》明确要求“健全因地制宜发展新质生产力体制机制”，并就推动产业深度转型升级作出重要部署。加快形成新质生产力，强化内生创新驱动力，改造升级传统产业，前瞻谋划新兴产业，超前布局建设未来产业，是我国构筑国家竞争新优势的核心动力源，也是抢占全球新一轮产业发展战略制高点的关键环节。
\par
    在全面高质量发展、提高新质生产力以及全球技术变革和竞争加剧的背景下，中国进行新一轮产业深度转型显得尤为必要。这一转型不仅是应对国内经济转型和提升国际竞争力的战略需求，也是应对全球产业变革和技术创新浪潮的必然选择。
\par
    从提升新质生产力的迫切需求来看，其不仅仅依赖于资本和劳动的传统投入，更依赖于技术创新、生产方式变革、资源的高效利用和产业的现代化。在中国经济进入“新常态”后，传统的依赖低成本劳动力和资源密集型产业的增长方式已经难以为继。提升新质生产力，尤其是通过技术创新推动生产效率和产业附加值的提升，已成为支撑经济高质量增长的核心动力。
\par
    从供给侧结构性改革的深化来看，这对实现经济高质量发展至关重要。产业深度转型是供给侧改革的核心内容之一，通过技术进步和产业升级，提升有效供给、缩小低效供给，能够更好地满足国内外市场的需求变化，推动经济增长由数量型向质量型转变。
\par
    产业结构的深度转型不仅需要推动传统产业的升级，还需要培育和发展新兴产业，如人工智能、新能源、高端装备制造等，这些新兴产业将成为经济增长的新引擎。中国需要加大对战略性新兴产业的支持，推动各行业间的融合创新，特别是推动制造业与互联网、人工智能、物联网、大数据等新兴技术的深度融合，以形成新的产业集群，推动新质生产力的快速增长。
\par
    从全球技术变革与产业竞争加剧来看，发达国家和其他技术领先国家正在加速技术创新并深度融合现代制造业和服务业。中国若不能及时适应这些变化，将面临在全球产业链中被边缘化的风险。因此，中国必须加强核心技术的自主研发，突破技术瓶颈，避免在全球产业竞争中陷入“卡脖子”问题。
\par
    以美国为代表的发达国家提出了“再工业化”战略，并加速制造业回流。与此同时，发展中国家特别是东南亚国家也在加大技术投资和产业升级，形成了对中国制造的强劲竞争压力。在这种全球产业竞争格局中，中国必须加速产业的高端化、智能化和绿色化转型，以提升产业附加值、增强技术竞争力，从而在全球市场中保持竞争优势。
\par
    从绿色发展与可持续转型来看，面对全球气候变化的挑战和国内日益严峻的环境压力，中国提出的“双碳”目标(碳达峰和碳中和)对中国的产业结构提出了深刻的转型要求。传统的重化工业模式，特别是高能耗、高排放的产业已不再适应可持续发展的需求。因此，中国必须加速推动产业的绿色低碳转型，在减少资源消耗、提高能效、促进绿色创新等方面取得突破，才能实现“双碳”目标，并引领全球绿色产业的未来。

\section{理论假说}
    假说1：产业转移推动技术进步，进而提升新质生产力。
\par
    产业转移通过吸引外部资源和技术，促进技术扩散与知识溢出，从而推动区域内技术水平的提升。这一过程中，高技术产业的引入和传统产业的改造升级，是实现新质生产力增长的关键途径。
\par
    假说2：产业结构优化通过引导资源向高新产业聚集，提升整体生产效率。
\par
    随着资源逐步从低附加值产业向高技术制造业和现代服务业转移，产业结构优化能够有效提高资源配置效率，进而带动全要素生产率的提升。这一作用机制在区域经济发展中尤为显著。
\par
    假说3：市场化改革、人才流动和创新机制的完善，将加速产业转移对新质生产力的推动作用。
\par
    市场化改革通过改善要素市场的运行效率，为产业转移提供制度保障；人才流动作为技术转移的重要载体，直接影响产业升级的速度和质量；创新机制的完善不仅提升了区域内的研发能力，还强化了高端产业的集聚效应，从而进一步扩大产业转移对新质生产力的正向作用。

\section{实证研究}
\subsection{研究路径和数据}
\subsubsection{实证研究路径}
    如前文所述，为了研究产业结构转型升级对新质生产力发展的影响，本文选取我国明确提出促进产业结构转型升级的里程碑事件作为分析对象，基准回归方程如下：
\begin{equation}
    \text{ln}nqp_{i,t} = \alpha + \beta_1 \text{ln}str_{i,t} + \beta_2 X_{i,t} + \mu_i + \eta_t + \varepsilon_{i,t}
\label{eq:1}
\end{equation}
\par    
    其中，$i$代表省份，$t$代表年份，$\mu _i$代表省份固定效应，$\eta _t$代表年份固定效应。$\text{ln}nqp_{i,t}$代表省份$i$在年份$t$新质生产力指标加1取对数，$\text{ln}str_{i,t}$表示省份$i$在年份$t$的产业结构升级指数加1取对数。$X_{i,t}$是省份层面的控制变量，为缓解控制变量和被解释变量之间的反向因果关系，方程中采用滞后一期的控制变量。
\par
方程\ref{eq:1}能够刻画产业结构转型升级对新质生产力发展的影响，但新质生产力水平本身具有丰富内涵和多种表现形式，为了估计新质生产力提高的全要素生产率，利用方程\ref{eq:2}进行回归分析：
\begin{equation}
    tfp_{i,t}=\alpha +\beta _1nqp_{i,t}+\beta _2struc_{i,t}+\beta _3X_{i,t}+\mu _i+\eta _t+\varepsilon _{i,t}
\label{eq:2}
\end{equation}
\par
    $tfp_{i,t}$代表省份$i$在年份$t$测算出的全要素生产率。
\subsubsection{实证研究数据以及指标的构建}
\subsubsubsection{省级经济数据}
\indent
    本文的省级数据主要来源于国家统计局，历年的《中国统计年鉴》以及各省统计年鉴和统计公报等，还来源于中国银行业微观市场结构测算与分析(刘岩等，2024)。本文未包含港、澳、台等数据样本，选取了中国大陆31个省份2003—2022年的数据。
\subsubsubsection{产业结构升级指数的构建}
\indent
    本文基于产业产值变动视角，探讨其对区域产业结构高级化与新质生产力增长的影响。在研究方法上，借鉴配第一克拉克定理，通过计算第二、第三产业产值占地区生产总值(GDP)的比重，测度区域产业结构发展水平。这一方法能够有效捕捉新兴产业、高端技术制造业及服务业在产业结构转型升级中的关键作用。具体公式如下：
\begin{equation}
    struc_{i,t} = \frac{Y_{2i,t} + Y_{3i,t}}{Y_{i,t}}
\label{eq:3}
\end{equation}
\par
    其中，$struc_{i,t}$ 表示 $i$ 省份 $t$ 年度的产业结构升级指数；$Y_{2i,t}$ 和 $Y_{3i,t}$ 分别代表 $i$ 省份第二、第三产业在 $t$ 年度的产业产值；$Y_{i,t}$ 表示 $i$ 省份 $t$ 年度的 GDP。由式\ref{eq:3}可知，随着产业结构从传统农业向工业和服务业转型，指标值呈上升趋势，表明区域产业结构逐步实现高级化转型升级。
\par
    根据指数数据，可以把2003—2022年中国产业结构高级化的特征和变化趋势概述如下：
\par
    首先，产业结构高级化程度逐年提升，省际差异逐渐收敛。从时间维度来看，2003—2022年中国产业结构高级化程度逐年攀升，指标均值从2003年的0.8512增长至2022年的0.9067，区间涨幅5.55\%，中位数则由0.8322逐步攀升至0.9099，区间涨幅高达7.77\%。
\par
2003—2015年产业结构高级化指数的中位数小于平均值，且从2007年开始，中位数逐渐缩小与均值之间的差距，反映出2003—2015年间中国31个省份产业高级化程度差异较大，趋向于偏态分布。
\begin{figure}[ht]
    \centering
    \includegraphics[scale=0.35]{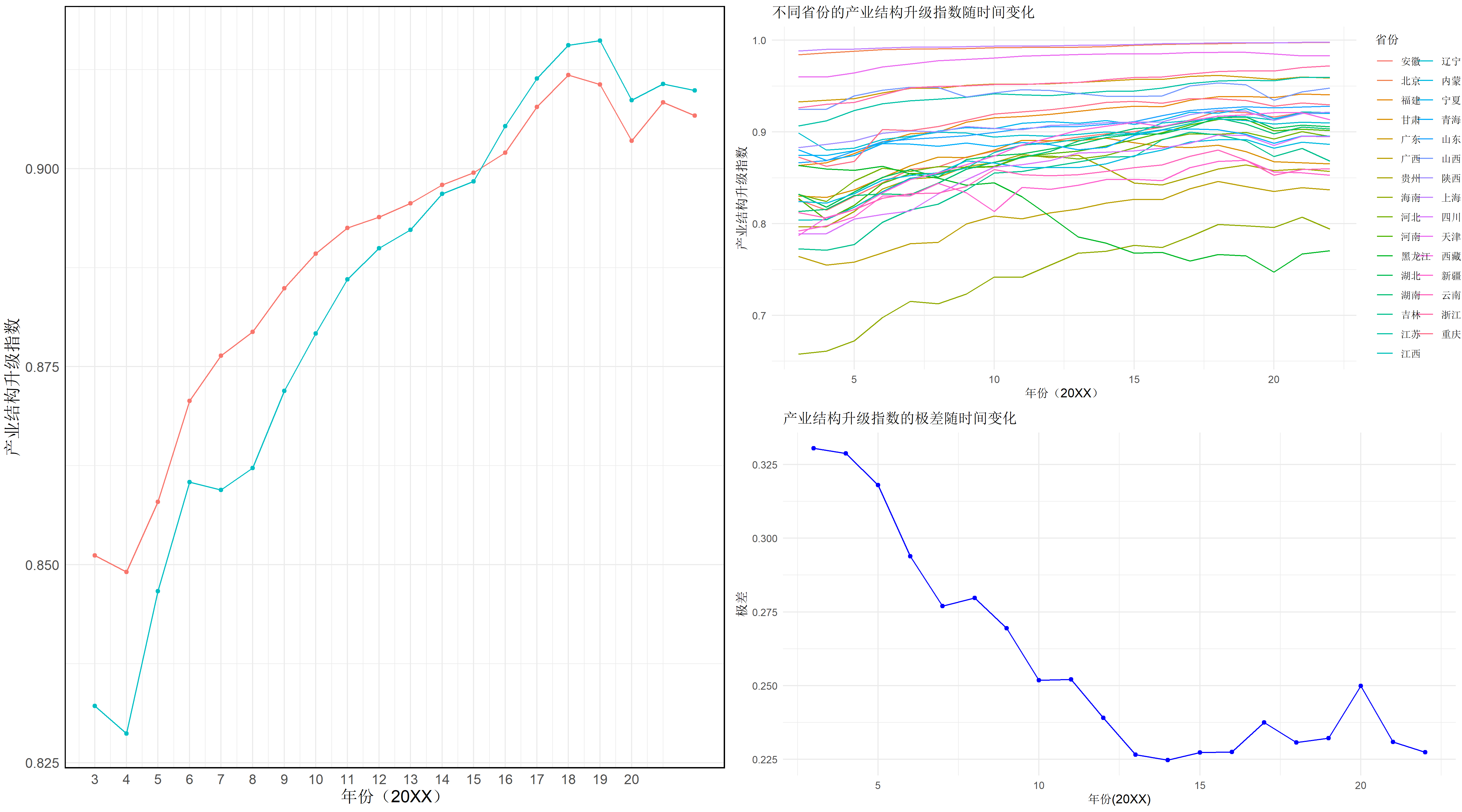}
    \caption{2003—2022年产业结构高级化水平}
    \label{fig0}
\end{figure}
\par
    如图\ref{fig0}所示，2003年至2022年，中国产业结构高级化水平整体呈显著提升态势，指标均值由2003年的较低水平攀升至2020年的高位区间，区间涨幅显著。同时，中位数从低位逐步上升，并于2016年首次超过均值，反映出省际差异逐渐缩小，产业结构水平趋于正态分布。然而，2020年受新冠疫情冲击，产业结构水平略有回落，但整体仍处于高分位区间，展现出中国经济的韧性。
\par
    从区域分布来看，产业结构高级化水平呈现出“东高西低”的梯度分布。数据显示，东部沿海地区尤其是直辖市(如上海、北京)在指标均值和中位数方面均处于领先地位；相比之下，中西部地区如海南、广西等省份产业结构水平较低，但仍具有较大的优化空间。
\par
    东部沿海地区产业结构升级指数的最大值接近1.0，反映出长三角、珠三角和京津冀城市群已形成以高端技术制造业和服务业为主的产业结构。在中西部地区，2003年至2013年间，产业结构升级指数的最小值快速上升，极差逐渐收敛。2014年起，最小值趋于稳定，极差维持在低位，显示出区域间产业结构升级步伐逐步趋同。
\subsubsubsection{新质生产力指标体系的构建}
\indent
    本文采用基于新质生产力的理论内涵构建的新质生产力的指标体系\cite{SLJY202406001}，通过新劳动者、新劳动资料、新劳动对象三项实体性要素和新技术、生产组织、数据要素三项渗透性要素使用客观赋权法中的熵值法对各个指标进行赋权：首先对各个指标进行无量纲化处理，其次计算指标比重、信息熵的信息熵冗余度，然后计算指标权重，最后加权即求出总指数综合得分(王军等，2021)。
\begin{figure}[ht]
    \centering
    \includegraphics[scale=0.35]{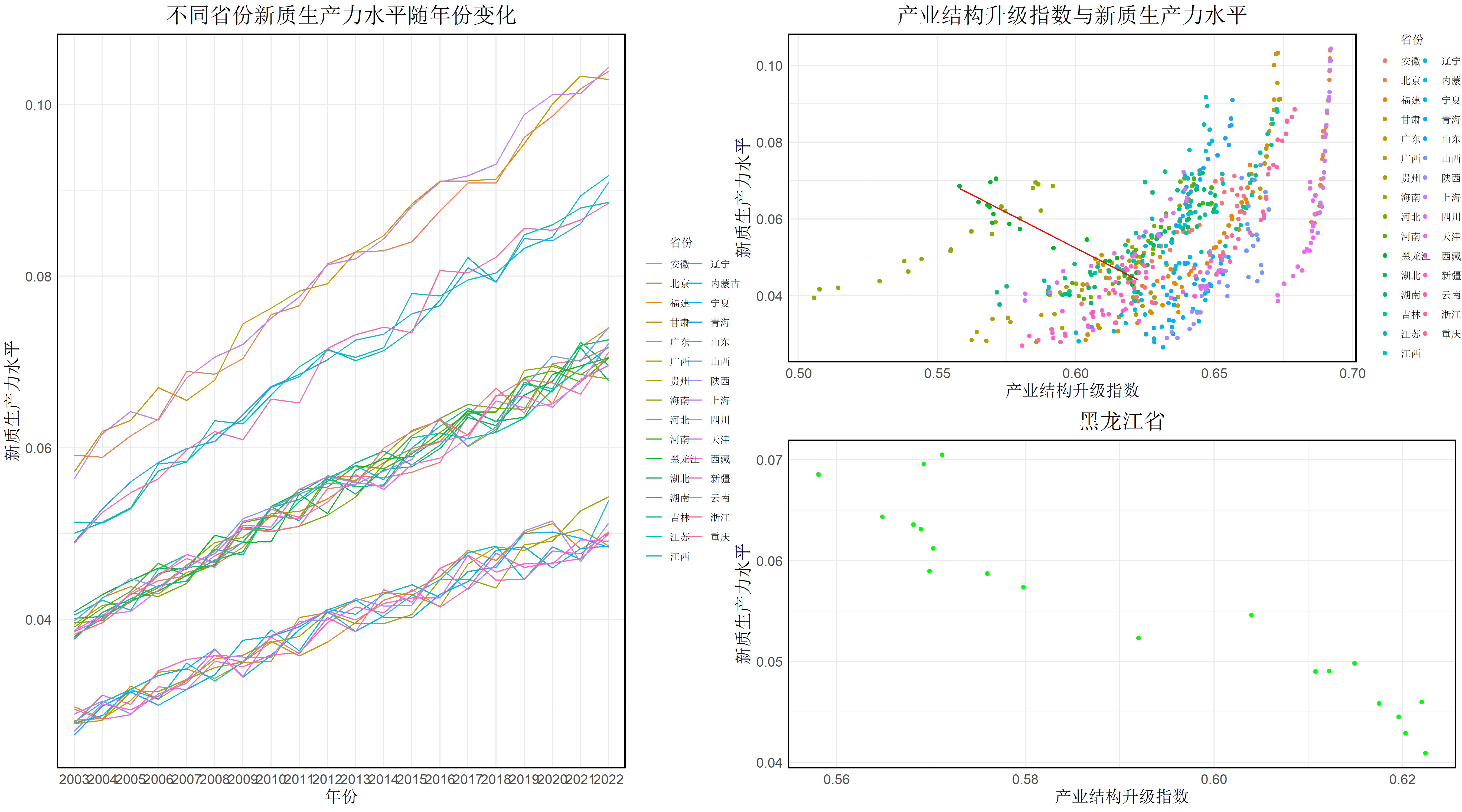}
    \caption{2003—2022年新质生产力水平}
    \label{fig1}
\end{figure}
\par
    测算结果如图\ref{fig1}所示，从测算结果来看，各省的新质生产力水平总体呈上升趋势，但由于个省份的发展条件不同，展现出明显的地区分别。上海、北京、广东新质生产力水平显著高于其他地区，山东、江苏、浙江、辽宁其次，云南、内蒙古、宁夏、广西、新疆、甘肃、西藏、贵州、陕西、青海处于最后梯队。其中最值得注意的是，黑龙江省是唯一一个新质生产力水平随产业结构高级化呈下降趋势的省份，这可能与该省的产业结构调整不力有关。
\subsubsubsection{全要素生产率指标的测算}
\indent
    本文采用基于总量细分行业面板数据测算的全要素生产率，从资本和劳动回归弹性系数，年均总量，总量的趋势变化和动态变化、OP协方差及其变化等方面测算的不同模型的结果。研究表明，宏观分行业面板数据更适用于非参数的数据包络分析—Malmquist指数测算方法，原因在于，参数方法中对行业之间生产技术相同的假设并不合理\cite{JJXU202102004}。
\begin{figure}[ht]
\centering
    \begin{subfigure}[b]{0.45\textwidth}
        \includegraphics[width=\textwidth]{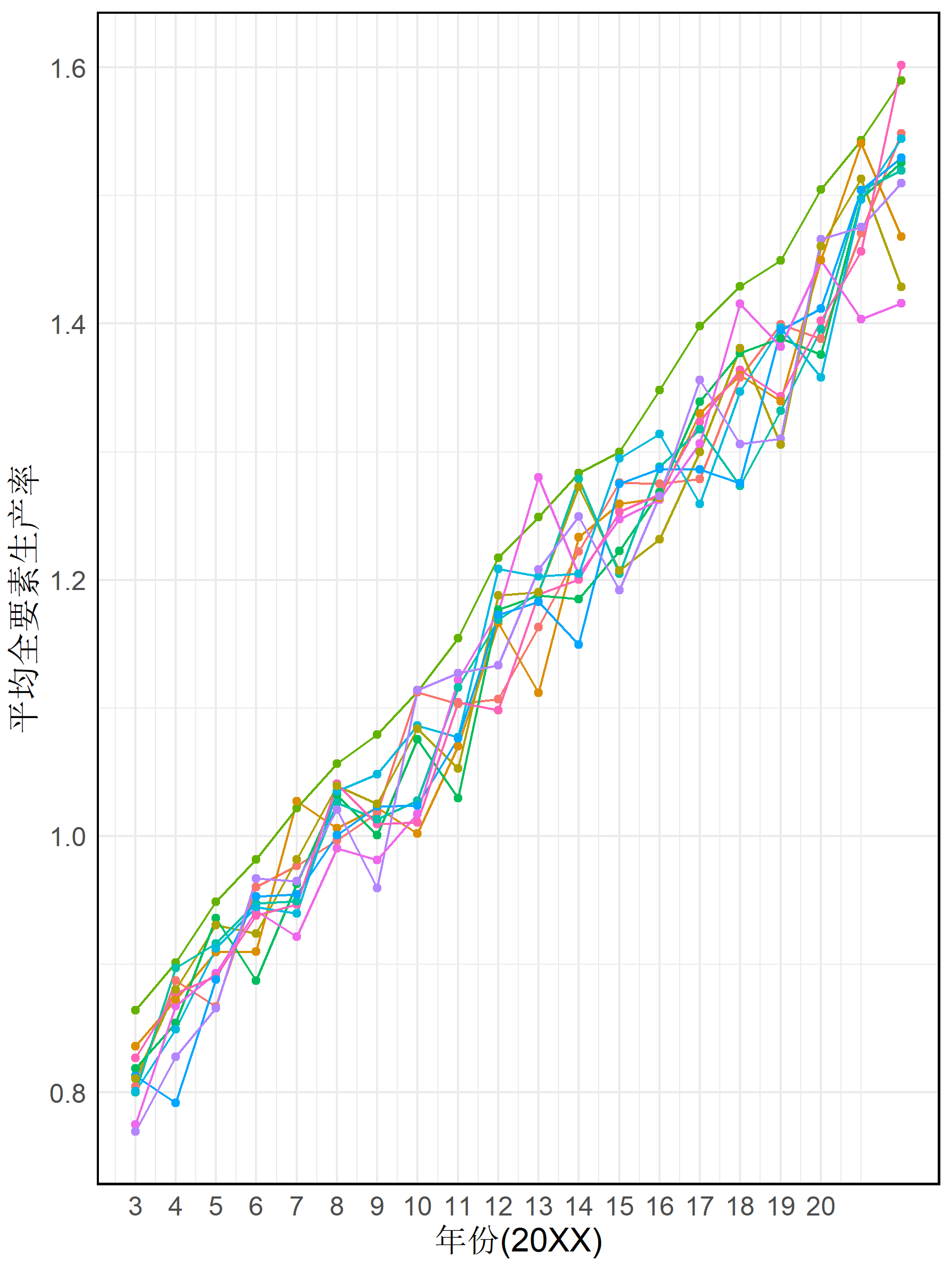}
        \caption{按省份平均}
        \label{fig2_1}
    \end{subfigure}
    \begin{subfigure}[b]{0.45\textwidth}
        \includegraphics[width=\textwidth]{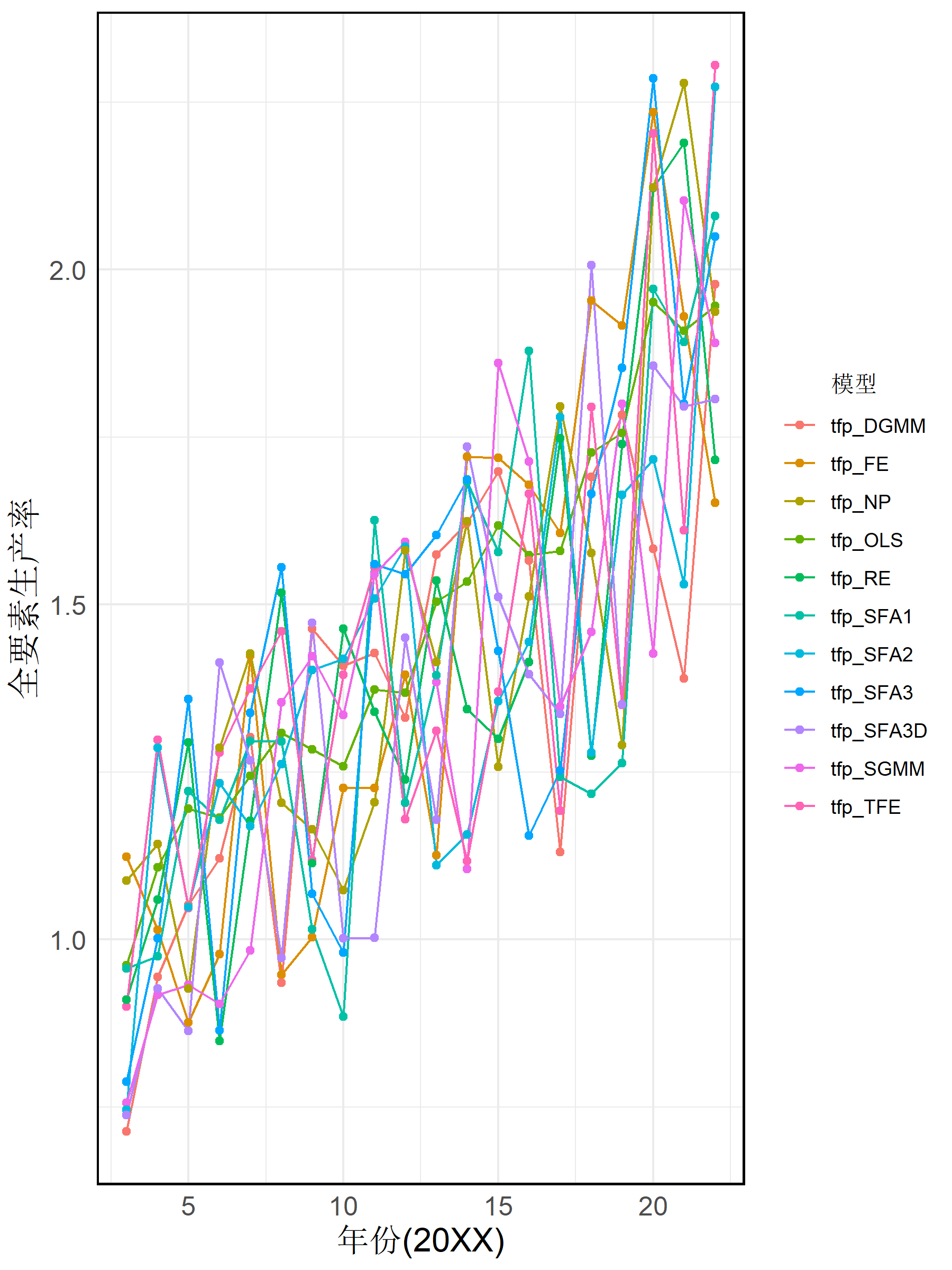}
        \caption{黑龙江省}
        \label{fig2_2}
    \end{subfigure}
\caption{不同全要素生产率测算模型的结果}
\label{fig2}
\end{figure}
\par
    图\ref{fig2}中图\ref{fig2_1}展示了不同的全要素生产率测算模型每年度的均值随时间的变化，从测算结果来看，总体上不同模型的全要素生产率总体呈上升趋势。图\ref{fig2_2}则展示了黑龙江省各种测算方法测算的全要素生产率随时间的变化。 
\subsubsubsection{控制变量}
\indent
    回归方程中的控制变量包括：(1)政府财政能力($fiscap_{it}$)采用地方财政支出与地方财政收入的比值表示；(2)政府研发投入($rd_{it}$)采用地方财政科学技术支出表示；(3)劳动力水平($labor_{it}$)采用城镇单位就业人员数量表示；(4)地方金融支持($finsup_{it}$)采用地方金融机构年末贷款余额表示；(5)教育密度($edudens_{it}$)采用地方财政教育支出与城区面积的比值表示；(6)城市化率($urb_{it}$)采用地方城镇人口与总人口的比值表示；(7)基础设施建设水平($infcon_{it}$)采用地区公路里程和地区铁路运营里程加和后与城区面积的比值表示(8)开放程度($open_{it}$)采用地区进出口总额(美元)表示。同时对各类比例变量都加一后取对数值。

\subsection{基准回归结果}
\subsubsection{基准回归}
    方程\ref{eq:1}的回归结果见表\ref{table2}。在表\ref{table2}第(1)—(3)列中逐步加入控制变量，$struc$的系数始终为正，但在控制变量仅加入地区基础发展水平时并不显著。回归结果表明，当一个地区的产业结构越高级，该地区的新质生产力水平会出现显著的增长。
\begin{table}[htbp]
    \renewcommand{\arraystretch}{1.25}
    \zihao{5}
    \caption{产业深度转型升级对新质生产力水平的影响}
    \centering
    \begin{tabularx}{\textwidth}{c c c c}
    \toprule
    \textbf{变量} & \textbf{(1)} & \textbf{(2)} & \textbf{(3)} \\ \midrule
    \textbf{$\ln(struc)$}  & 0.0095  & 0.0155**  & 0.0208***    \\
                               & (0.0066) & (0.0060) & (0.0065)   \\
    控制变量      & 基础发展水平 & 基础发展水平；政府能力 & 基础发展水平；政府能力；经济活跃度 \\
    时间固定效应 & Y & Y & Y \\
    个体固定效应 & Y & Y & Y \\
    \text{$R^2$} & 0.4276 & 0.4792 & 0.4955 \\
    样本量 & 558 & 558 & 558 \\ \bottomrule
    \end{tabularx}
    \begin{flushleft}
    \zihao{6} 注：括号中为回归标准误，*、**和***分别表示在10\%、5\%和1\%的水平上显著。以下各表同。
    \end{flushleft}
\label{table2}    
\end{table}
\par
    表\ref{table3}为方程的稳健性检验。考虑到如图所示黑龙江省数据的异常表现，表\ref{table3}第(1)列是剔除黑龙江省的样本。考虑到直辖市在政治与经济上的特殊性，表\ref{table3}第(2)列是去掉4个直辖市的样本。同样的，表\ref{table3}第(3)列是去掉5个少数民族自治区的样本。三种稳健性检验的结果均表明，产业深度转型升级对新质生产力水平有显著的促进作用。
\begin{table}[htbp]
    \renewcommand{\arraystretch}{1.25}
    \zihao{5}
    \caption{基准回归的稳健性检验}
    \centering
    \begin{tabularx}{\textwidth}{c >{\centering\arraybackslash}X >{\centering\arraybackslash}X >{\centering\arraybackslash}X} 
    \toprule
    \textbf{变量} & \textbf{(1)} & \textbf{(2)} & \textbf{(3)} \\
                & \textbf{去掉黑龙江省} & \textbf{去掉直辖市} & \textbf{去掉自治区} \\ \midrule
    \textbf{$\ln(struc)$}  & 0.0211***  & 0.0235***  & 0.0263*** \\
                            & (0.0079) & (0.0063) & (0.0060) \\
    控制变量     & Y & Y & Y \\
    时间固定效应 & Y & Y & Y \\
    个体固定效应 & Y & Y & Y \\
    \text{$R^2$} & 0.5022 & 0.4342 & 0.5025 \\
    \text{$N$} & 540 & 486 & 468 \\ \bottomrule
    \end{tabularx}
\label{table3}    
\end{table}
\subsubsection{内生性问题}
    考虑到解释变量与被解释变量之间可能存在内生性，本文进一步引入被解释变量的滞后一期项作为工具变量并检验新质生产力水平是否存在一定的路径依赖。分别使用两阶段最小二乘、差分GMM以及系统GMM三种不同的方法进行了估计。在样本量为大N小T的情况下，利用基于GMM的工具变量估计法往往更加有效。在存在弱工具变量问题时，系统GMM估计更有效。构建动态面板模型如下：
\begin{equation}
\text{ln}nqp_{i,t}=\alpha +Z_1\text{ln}nqp_{i,1}+\beta _1\text{ln}str_{i,t}+\beta _2X_{i,t}+\mu _i+\eta _t+\varepsilon _{i,t}
\label{eq:4}
\end{equation}
\par
    方程\ref{eq:4}的回归结果见表\ref{table4}。表\ref{table4}第(1)列是所有省份的新质生产力水平数据作为被解释变量的两阶段最小二乘法回归结果，第(2)列是差分GMM法的回归结果，第(3)列是系统GMM法的回归结果。第(4)—(6)列是将去除黑龙江省后的新质生产力水平数据作为被解释变量分别进行两阶段最小二乘、差分GMM和系统GMM的回归结果。从核心解释变量的系数看，可以发现在结果显著的情况下产业结构升级指数对新质生产力水平的影响均为正，且两阶段最小二乘法的回归结果较好。从具体的回归结果看，两阶段最小二乘法在全样本下的估计结果显示产业结构升级指数的系数在5\%的统计水平上显著为正，这表明2003年以来的结构变动在整体上有助于促进当地的新质生产力水平发展。产业结构升级对高质量发展存在“结构性红利”，通过促进地区的资源合理配置进而推动新质生产力水平发展。两阶段最小二乘、差分GMM和系统GMM在去除直辖市样本下的估计结果均为正且较为显著，这表明产业结构升级指数对新质生产力水平的作用因为不同的地区的资源禀赋、发展条件不同，存在一定的地区分别。
\begin{table}[htbp]
    \renewcommand{\arraystretch}{1.25}
    \zihao{5}
    \caption{引入工具变量后动态面板的回归结果}
    \centering
    \begin{tabularx}{\textwidth}{c >{\centering\arraybackslash}X >{\centering\arraybackslash}X >{\centering\arraybackslash}X >{\centering\arraybackslash}X >{\centering\arraybackslash}X >{\centering\arraybackslash}X} 
    \toprule
    \textbf{变量} & \textbf{(1)} & \textbf{(2)} & \textbf{(3)} & \textbf{(4)} & \textbf{(5)} & \textbf{(6)}\\
                & \textbf{2SLS} & \textbf{差分GMM} & \textbf{系统GMM} & \textbf{2SLS} & \textbf{差分GMM} & \textbf{系统GMM} \\
                & \textbf{全样本} & \textbf{全样本} & \textbf{全样本} & \textbf{去掉直辖市} & \textbf{去掉直辖市} & \textbf{去掉直辖市} \\\midrule
    \textbf{$\ln(struc)$}  & 0.8103**  &0.0048  &0.0123 & 0.5706***  & 0.0043***  & 0.0114* \\ 
                            & (0.3162) & (0.0045) & (0.0394) & (0.2046)  & (0.0004)  & (0.0050) \\
    控制变量     & Y & Y & Y & Y & Y & Y \\
    时间固定效应 & Y & Y & Y & Y & Y & Y \\
    个体固定效应 & Y & Y & Y & Y & Y & Y \\
    \text{$R^2$} & 0.8523 &  &  & 0.9027 &  &  \\
    \text{$N$} & 558 & 558 & 558 & 486 & 486 & 486 \\ \bottomrule
    \end{tabularx}
\label{table4}    
\end{table}
\subsubsection{不同省份的异质性分析}
    为了分析产业结构深度转型升级对不同发展水平的省份的影响，本文将回归方程\ref{eq:1}中的被解释变量替换为不同发展类型的省份，将省份按照图\ref{fig1}中新质生产力水平的差异情况分为由高到低四个梯队，普通发展省份为位于第二、三和四梯队的省份，中低发展省份为位于三、四梯队的省份。回归结果见表\ref{table5}，表\ref{table5}的第(1)—(3)列的被解释变量为普通发展省份的数据，即去除上海、北京和广州之后的数据，表的第(4)—(6)列的被解释变量为中低发展省份的数据，并分别用混合面板回归、两阶段最小二乘法与差分GMM进行回归分析。
\begin{table}[htbp]
    \renewcommand{\arraystretch}{1.25}
    \zihao{5}
    \caption{产业深度转型升级对不同情况省份的影响}
    \centering
    \begin{tabularx}{\textwidth}{c >{\centering\arraybackslash}X >{\centering\arraybackslash}X >{\centering\arraybackslash}X >{\centering\arraybackslash}X >{\centering\arraybackslash}X >{\centering\arraybackslash}X} 
    \toprule
    \textbf{变量} & \textbf{(1)} & \textbf{(2)} & \textbf{(3)} & \textbf{(4)} & \textbf{(5)} & \textbf{(6)}\\
                & \textbf{OLS} & \textbf{2SLS} & \textbf{差分GMM} & \textbf{OLS} & \textbf{2SLS} & \textbf{差分GMM} \\
                & \textbf{普通发展省份} & \textbf{普通发展省份} & \textbf{普通发展省份} & \textbf{中低发展省份} & \textbf{中低发展省份} & \textbf{中低发展省份} \\\midrule
    \textbf{$\ln(struc)$}  & 0.0229***  & 0.5668***  & 0.0056*** & 0.0192**  & 0.5219**  & 0.0013*** \\ 
                            & (0.0061) & (0.2085) & (0.0004) & (0.0083)  & (0.2278)  & (0.0002) \\
    控制变量     & Y & Y & Y & Y & Y & Y \\
    时间固定效应 & Y & Y & Y & Y & Y & Y \\
    个体固定效应 & Y & Y & Y & Y & Y & Y \\
    \text{$R^2$} & 0.3495 & 0.8871 &  & 0.2533 & 0.8574 &  \\
    \text{$N$} & 504 & 504 & 504 & 432 & 432 & 432 \\ \bottomrule
\end{tabularx}
\label{table5}    
\end{table}
\par
    如前所述，产业结构升级对高质量发展存在“结构性红利”，通过促进地区的资源合理配置进而推动新质生产力水平发展。可以从具体的回归结果看到，在不同方法下的回归结果均显示产业结构升级指数在普通发展省份的系数大于中低发展省份，表明产业结构升级对普通发展省份的经济增长与新质生产力提升具有更强的推动作用。这种差异可能源于普通发展省份在基础设施、技术储备以及人才资源等方面的相对优势，使其更能有效吸收和转化产业结构优化带来的红利。作者认为，中低发展省份尽管在产业结构升级中受到一定程度的制约，但通过政策扶持与资源倾斜，仍可以逐步缩小与普通发展省份的差距。例如，东部沿海省份通过加快新兴产业集群建设，在区域协调发展中起到了引领作用，而中西部地区则需要更多依赖政策支持来优化产业结构并提升新质生产力水平。

\subsection{新质生产力水平的发展：来自全要素生产率的证据}
    “新质生产力以全要素生产率大幅提升为核心标志，特点是创新，关键在质优，本质是先进生产力。”如前文所述，本文选取非参数的DEA方法与TFE方法测算出的全要素生产率通过方程\ref{eq:2}论证新质生产力水平的提升对全要素生产率具有正向的作用，并进行估计。方程\ref{eq:2}的回归结果见表\ref{table6}。
\begin{table}[htbp]
    \renewcommand{\arraystretch}{1.25}
    \zihao{5}
    \caption{新质生产力水平对全要素生产率的影响}
    \centering
    \begin{tabularx}{\textwidth}{c >{\centering\arraybackslash}X >{\centering\arraybackslash}X >{\centering\arraybackslash}X >{\centering\arraybackslash}X >{\centering\arraybackslash}X >{\centering\arraybackslash}X} 
    \toprule
    \textbf{变量} & \textbf{(1)} & \textbf{(2)} & \textbf{(3)} & \textbf{(4)} & \textbf{(5)} & \textbf{(6)}\\
                    & \textbf{OLS} & \textbf{差分GMM} & \textbf{系统GMM} & \textbf{OLS} & \textbf{差分GMM} & \textbf{系统GMM} \\
                    & \textbf{$tfp_{DEA}$} & \textbf{$tfp_{DEA}$} & \textbf{$tfp_{DEA}$} & \textbf{$tfp_{TFE}$} & \textbf{$tfp_{TFE}$} & \textbf{$tfp_{TFE}$} \\\midrule
    \textbf{$nqp$}  & 11.34***  & 0.3215*  & 0.0961*** & 6.9765 & 0.0921  & 0.0413 \\ 
                                & (3.7452) & (0.1251) & (0.0236) & (4.7306)  & (0.0581)  & (0.1946) \\
    控制变量     & Y & Y & Y & Y & Y & Y \\
    时间固定效应 & Y & Y & Y & Y & Y & Y \\
    个体固定效应 & Y & Y & Y & Y & Y & Y \\
    \text{$R^2$} & 0.0684 &  &  & 0.0662 &  &  \\
    \text{$N$} & 558 & 558 & 558 & 558 & 558 & 558 \\ \bottomrule
\end{tabularx}
\label{table6}    
\end{table}
\par
    可以看到，对非参数的DEA方法测算出来的全要素生产率在统计学上均显著，而对TFE方法测算出来的全要素生产率在统计学上不显著，但都系数都为正。这表明新质生产力水平对全要素生产率的提升具有显著的正向作用，但在不同的测算方法下，新质生产力水平对全要素生产率的影响存在一定的差异。因此，本文认为新质生产力水平的提升对全要素生产率的提升具有显著的正向作用，但在具体的测算方法下存在一定的差异。

\section{全面提升新质生产力的未来发展之路}
    全球各国在产业发展过程中都经历了不同的转型阶段。美国、德国和日本等工业化较早的国家，都经历过从传统农业经济到制造业驱动经济的历史进程。在20世纪中后期，随着信息技术的飞速发展，尤其是互联网和人工智能的应用，发达国家纷纷进入了产业结构优化升级的阶段，开始大力推动制造业向高端化、智能化、绿色化发展。其产业转型升级的过程、特征与驱动要素具有重要的借鉴意义。
\par
    虽然进行去工业化调整是资本主义从工业社会向后工业社会转变时的必然结果，但是与英美等国依靠市场进行产业转移式的去工业化不同，德国与日本更加偏重以政府主导为主的产业升级式工业化发展。即通过国家政策对产业结构调整进行干预，推动制造业增长方式从传统要素驱动向科技创新驱动转变。在此过程中，德国与日本的工业化表现，既具有产业转移型的“去工业化”特征，也具有产业升级型的“再工业化”特征。中国与德国日本相似，产业发展历程呈现出明显的阶段性特征。从改革开放以来，中国经济依赖于低成本劳动力和大量投资的粗放型增长方式，发展成为世界制造业中心。党的十八大以来，中国正在成为世界高端制造业的中心。
\par
    在20世纪70年代，美国经历了制造业的衰退，特别是在传统工业区出现了严重的产业空心化现象。这主要归因于全球化带来的制造业外迁、技术进步导致的自动化，以及国内外竞争的加剧。美国政府采取了再工业化战略，推动高科技产业和创新经济的发展。产业转型过程中出现了就业岗位减少、技能不匹配等问题。德国在推进“工业4.0”战略过程中，面临着传统制造业向数字化、智能化转型的挑战。这一过程中，企业需要大量投资于新技术和员工培训，以适应新的生产模式。日本在战后迅速实现了工业化，在产业结构升级过程中，日本面临了人口老龄化、内需不足、技术创新乏力等挑战。过度依赖出口导向型经济，使得日本在全球经济波动中受到较大影响。在新一轮全球再工业化浪潮下，未来中国需要吸取德国与日本的工业化经验，不断加大研发支出，推进区域经济合作深入发展。最终实现“保持制造业比重基本稳定，巩固壮大实体经济根基”的十四五发展目标。
\par
    习总书记指出：“科技成果转化为现实生产力，表现形式为催生新产业、推动产业深度转型升级。”产业是生产力的载体。加快发展新质生产力，必须加快建设现代化产业体系。中国的产业深度转型升级是全面提升新质生产力、应对全球竞争挑战、实现经济高质量发展的关键所在。通过加强技术创新、推动绿色低碳发展、提升全球产业链地位和深化供给侧改革，中国有能力实现产业结构的深度升级，促进新质生产力的全面提升。在新的全球发展格局下，全面提升新质生产力不仅是中国经济未来发展的必由之路，也是中国在全球经济中占据更强竞争力的关键。

\section{结语}
    产业是生产力变革的具体表现形式。本文回顾了中国产业结构演变的三阶段历程，从改革开放时我国纠正失衡和均衡发展的产业结构到20世纪末在重化工业重启下以第二产业为主导的产业结构，最后到党的十八大后以第三产业为主导的经济新常态下的产业结构，尤其着重探讨了2020年以来中国经济进入高质量发展的新时代下的产业结构深度升级与发展。并在全面高质量发展、提高新质生产力以及全球竞争力的背景下，从提升新质生产力的迫切需求、供给侧结构性改革、全球技术变革与竞争加剧、绿色发展与可持续化转型四个方面详细论述了中国进行新一轮产业深度转型的必要性。
\par
    本文构建了考察产业结构深度转型的产业结构升级指数，通过实证研究考察了我国产业结构深度转型对新质生产力发展的促进作用。研究结果表明，在省级层面，产业深度转型升级能够显著促进新质生产力水平发展，但存在较为明显的地区差异。同时，本文还研究了新质生产力与全要素生产率之间的关系。结论表明，在不同的测算方法下，新质生产力水平的提升对全要素生产率的提升有显著正向作用，证明了习总书记在主持中共中央政治局第十一次集体学习时提出的新质生产力水平提升的核心标志是全要素生产率的提升。
\par
    最后，本文梳理了世界各国产业结构转型升级的历程与曾经面临的挑战，并结合我国的实际情况与实证研究结果分析探讨了我国未来在产业结构深度转型升级下全面提升新质生产力的未来发展之路。

\nocite{*}
\printbibliography

@article{10.1257/aer.20230133,
  author  = {Fujiwara, Ippei and Matsuyama, Kiminori},
  title   = {A Technology-Gap Model of 'Premature' Deindustrialization},
  journal = {American Economic Review},
  volume  = {114},
  number  = {11},
  year    = {2024},
  month   = {November},
  pages   = {3714–45},
  doi     = {10.1257/aer.20230133},
  url     = {https://www.aeaweb.org/articles?id=10.1257/aer.20230133}
}

@article{a9f04cf5-4dfa-3113-890c-9a9d325056f8,
  issn      = {13814338, 15737020},
  url       = {https://www.jstor.org/stable/48700561},
  author    = {Dani Rodrik},
  journal   = {Journal of Economic Growth},
  number    = {1},
  pages     = {pp. 1--33},
  publisher = {Springer},
  title     = {Premature deindustrialization},
  urldate   = {2024-11-25},
  volume    = {21},
  year      = {2016}
}

@article{FJLW202009009,
  author  = {史丹 and 李鹏 and 许明},
  title   = {产业结构转型升级与经济高质量发展},
  journal = {福建论坛(人文社会科学版)},
  volume  = {},
  number  = {09},
  pages   = {108-118},
  year    = {2020},
  issn    = {1671-8402}
}

@article{GGYY201512007,
  author  = {于斌斌},
  title   = {产业结构调整与生产率提升的经济增长效应——基于中国城市动态空间面板模型的分析},
  journal = {中国工业经济},
  volume  = {},
  number  = {12},
  pages   = {83-98},
  year    = {2015},
  issn    = {1006-480X},
  doi     = {10.19581/j.cnki.ciejournal.2015.12.007}
}

@article{GILSING2011638,
  title    = {Differences in technology transfer between science-based and development-based industries: Transfer mechanisms and barriers},
  journal  = {Technovation},
  volume   = {31},
  number   = {12},
  pages    = {638-647},
  year     = {2011},
  issn     = {0166-4972},
  doi      = {https://doi.org/10.1016/j.technovation.2011.06.009},
  url      = {https://www.sciencedirect.com/science/article/pii/S0166497211000927},
  author   = {Victor Gilsing and Rudi Bekkers and Isabel Maria {Bodas Freitas} and Marianne {van der Steen}}
}

@article{GJMY202407002,
  author  = {洪俊杰 and 陈洋 and 杨志浩},
  title   = {中国产业转移的战略考量:特征、动因与政策展望},
  journal = {国际贸易},
  volume  = {},
  number  = {07},
  pages   = {11-21},
  year    = {2024},
  issn    = {1002-4999},
  doi     = {10.14114/j.cnki.itrade.2024.07.006}
}

@article{https://doi.org/10.2307/1235188,
  author  = {Goodman, Bernard},
  title   = {The Strategy of Economic Development, Albert O. Hirschman. (Yale Studies in Economics: 10.) New Haven: Yale University Press, 1958. Pp. xiii, 217. \$4.50},
  journal = {American Journal of Agricultural Economics},
  volume  = {41},
  number  = {2},
  pages   = {468-469},
  doi     = {https://doi.org/10.2307/1235188},
  url     = {https://onlinelibrary.wiley.com/doi/abs/10.2307/1235188},
  eprint  = {https://onlinelibrary.wiley.com/doi/pdf/10.2307/1235188},
  year    = {1959}
}

@article{JHLT20240102,
  author  = {方服前 and 付琦},
  title   = {产业结构升级的经济增长效应——基于2000-2020年中国31个省份面板数据的实证分析},
  journal = {江汉论坛},
  volume  = {},
  number  = {01},
  pages   = {12-25},
  year    = {2024}
}

@article{JJDL201605014,
  author  = {李献波 and 林雄斌 and 孙东琪},
  title   = {中国区域产业结构变动对经济增长的影响},
  journal = {经济地理},
  volume  = {36},
  number  = {05},
  pages   = {100-106},
  year    = {2016},
  issn    = {1000-8462},
  doi     = {10.15957/j.cnki.jjdl.2016.05.014}
}

@article{JJXU202102004,
  author  = {田友春 and 卢盛荣 and 李文溥},
  title   = {中国全要素生产率增长率的变化及提升途径——基于产业视角},
  journal = {经济学(季刊)},
  volume  = {21},
  number  = {02},
  pages   = {445-464},
  year    = {2021},
  issn    = {2095-1086},
  doi     = {10.13821/j.cnki.ceq.2021.02.04}
}

@article{JJYJ201811009,
  author  = {周茂 and 陆毅 and 李雨浓},
  title   = {地区产业升级与劳动收入份额:基于合成工具变量的估计},
  journal = {经济研究},
  volume  = {53},
  number  = {11},
  pages   = {132-147},
  year    = {2018},
  issn    = {0577-9154}
}

@article{JJYJ202409001,
  author  = {罗知 and 刘岩 and 陆利平 and 刘成},
  title   = {银行业对外开放与内资银行竞争和发展——兼论统筹银行业开放与安全},
  journal = {经济研究},
  volume  = {59},
  number  = {09},
  pages   = {4-22},
  year    = {2024},
  issn    = {0577-9154}
}

@article{SLJY201712002,
  author  = {田友春 and 卢盛荣 and 靳来群},
  title   = {方法、数据与全要素生产率测算差异},
  journal = {数量经济技术经济研究},
  volume  = {34},
  number  = {12},
  pages   = {22-40},
  year    = {2017},
  issn    = {1000-3894},
  doi     = {10.13653/j.cnki.jqte.2017.12.002}
}

@article{SLJY202406001,
  author  = {韩文龙 and 张瑞生 and 赵峰},
  title   = {新质生产力水平测算与中国经济增长新动能},
  journal = {数量经济技术经济研究},
  volume  = {41},
  number  = {06},
  pages   = {5-25},
  year    = {2024},
  issn    = {1000-3894},
  doi     = {10.13653/j.cnki.jqte.20240418.001}
}

@article{ZJSY201801014,
  author  = {郭旭红 and 武力},
  title   = {新中国产业结构演变述论(1949—2016)},
  journal = {中国经济史研究},
  volume  = {},
  number  = {01},
  pages   = {133-142},
  year    = {2018},
  issn    = {1002-8005}
}

\end{CJK}
\end{document}